\documentstyle[aps,12pt]{revtex}

\begin{document}
\title{Dirac Quantization of the Chern--Simons Field Theory in the Coulomb
Gauge}
\author{F. Ferrari\thanks{The work of F. Ferrari
has been supported in part by the European Union, TMR Programme, under
grant ERB4001GT951315}
$^{a,b}$ and I. Lazzizzera $^{a,c}$}
\address{
$^a$ Dipartimento di Fisica, Universit\'a di Trento, 38050 Povo (TN),
Italy\\
$^b$ LPTHE \thanks{%
Laboratoire associ\'e No. 280 au CNRS},
{\it Universit\'e Pierre er Marie Curie--PARIS VI},
{\it Universit\'e Denis Diderot--Paris VII},
{\it Boite $126$, Tour 16, $1^{er}$ \'etage},
{\it 4 place Jussieu},
{\it F-75252 Paris CEDEX 05, FRANCE}\\
$^c$ INFN, Gruppo Collegato di Trento, Italy}
\date{September 1996}
\maketitle
\vspace{-3.5in} \hfill{Preprint PAR-LPTHE 96-38, UTF 386/96} \vspace{3.4in}
\begin{abstract}
In this letter the
Chern--Simons field theories are studied in the Coulomb gauge
using the Dirac's canonical formalism for constrained
systems.
As a strategy,
we first work out the constraints and then quantize, replacing the
Dirac brackets with quantum commutators. We find that the
Chern--Simons field theories become two
dimensional models with no propagation along the time direction.
Moreover, we prove that,
despite of the presence of non-trivial self-interactions in the
gauge fixed functional, the commutation relations between the fields
are trivial at any order in perturbation theory
in the absence of couplings with matter fields.
If these couplings are present, instead, the commutation relations become
rather
involved, but it is still possible to study their main properties and to show
that they vanish at the tree level.
\end{abstract}

\section{Introduction}
In this letter we investigate the Chern--Simons (C--S) field
theories \cite{csft,csftt,witten}
with gauge group $SU(n)$ in the Coulomb gauge
using the Dirac's formalism for constrained
systems \cite{dirac,hrt}.
As it happens in the case of the more popular covariant gauges,
also in this gauge the C-S functional contains self-interactions,
but the Feynman rules simplify considerably and can be explicitly
derived  even on
space-times with a non--flat spatial section \cite{ffprd}.
Another advantage of the Coulomb gauge is that there are no
time derivatives in the gauge fixed action, so that the
C--S theory becomes
in practice a two-dimensional model.
Despite of
many physical and mathematical applications of the
C--S  field theories \cite{witten,csappl}, however,
until now only a few calculations
have been performed in the Coulomb gauge \cite{ffprd,cgincs,gg}.
One of the main reasons is that,
already in the case of the Yang--Mills field theories,
several perplexities arise concerning the use of this
gauge \cite{taylor,leibbrandt,chetsa}.
Analogous problems are unfortunately present also in C--S field theories,
but in a milder form, so that 
these models
provide an important laboratory in order to study the possible remedies.
For example, in the abelian case it is known that
the so-called Maxwell--Chern--Simons (MCS) theory is affected by
the presence of infrared divergences in the Coulomb gauge
\cite{csft,csftt}. Nevertheless, it has recently been shown 
in ref. \cite{gg} that the theory can be consistently
worked out and that for instance the M\"oller scattering
amplitudes computed in the Coulomb gauge and in the
covariant gauges coincide at all orders in perturbation theory.
Other tests confirming the safety of the Coulomb gauge in the MCS
models can also be found in \cite{gg}.

On the other side,
the ambiguities in the Yang-Mills Feynman integrals
pointed out in \cite{taylor}
arise as well in the nonabelian C--S field theories due
to the absence of time derivatives
in the action \cite{ffprd}. A simple recipe to regularize such ambiguities
has been proposed and successfully tested
in first order calculations \cite{ffprd}.
However, a detailed investigation of the consistency of the
nonabelian C--S field
theories in the Coulomb gauge
at any perturbative order is still missing. As one of the steps
to fill this gap,
we exploit in this letter the formalism of
Dirac's canonical approach to constrained systems \cite{dirac,hrt}.

We notice that, besides some subtleties already noticed in
\cite{linni}, the derivation of the final Dirac brackets requires in the
Coulomb gauge some care with distributions.
Moreover, the final commutation
relations (CR's) between the fields obtained here
are rather involved. At a first sight, this is surprising in topological
field theories with vanishing
Hamiltonian and without degrees of freedom. However, at
least in the case considered here,
in which there are no interactions with matter fields,
we show that
this contradiction is only apparent. As a matter of fact,
taking into account the Gauss law
and the Coulomb gauge fixing, it can be proved that the commutation
relations between the gauge fields vanish identically
at any perturbative order as expected.
In this way we discover that the Chern--Simons field theories in the Coulomb
gauge are not only perturbatively finite as has been checked
in the covariant gauges \cite{covgau},
but also free.
This is  not a priori evident, because in the Coulomb gauge the 
C--S functional contains non--trivial self--interaction terms.

The material presented in this paper is divided as follows.
In the next Section we will present our results. In the Conclusions
we will discuss the open problems and the possible further developments.

\section{Canonical Quantization of the C--S Field Theory in the
Coulomb Gauge}
\vspace{1cm} The Lagrangian of the pure $SU(N)$ Chern--Simons (C--S) field
theory in three dimensions is given by 
\begin{equation}
L_{CS}=\frac s{8\pi }\epsilon ^{\mu \nu \rho }
\left( A_\mu ^a\partial _\nu A_\rho
^a-\frac 13f^{abc}A_\mu ^aA_\nu ^bA_\rho ^c\right)  \label{lagrangian}
\end{equation}
where $s$ is a dimensionless coupling constant and $A_\mu ^a$ is the gauge
potential. Greek letters $\mu ,\nu ,\rho ,\ldots =0,1,2$ will denote
space--time indices while the first latin letters $a,b,c,\ldots =1,\cdots ,N$
will denote color indices. Moreover, the totally antisymmetric tensor $%
\epsilon ^{\mu \nu \rho }$ is defined by the convention $\epsilon ^{012}=1$.
Finally, a Minkowski metric $g_{\mu \nu }=$diag$(1,-1,-1)$ will be assumed.
To derive the C--S Hamiltonian, we have to compute the canonical momenta: 
\begin{equation}
\pi ^{\mu ,a}\left( {\bf x},t\right) =\frac{\delta S_{CS}}{\delta \left(
\partial _0A_\mu \left( {\bf x},t\right) \right) }  \label{cmomdef}
\end{equation}
Here we have put $S_{CS}=\int d^3xL_{CS}$, $t=x^0$
and ${\bf x}=\left( x^1,x^2\right) $.

The nonvanishing
Poisson brackets (PB) among canonical variables are: 
\[
\left\{ A_\mu ^a\left( {\bf x},t\right) ,\pi _\nu ^b\left( {\bf y},t\right)
\right\} =\delta ^{ab}g_{\mu \nu }\delta ^{(2)}\left( {\bf x}-{\bf y}\right) 
\]
From eqs. (\ref{lagrangian}) and (\ref{cmomdef})
we obtain:
\begin{equation}
\pi ^{0,a} =0  \qquad\qquad\qquad
\pi ^{i,a} =\frac s{8\pi }\epsilon ^{ij}A_j^a  \label{canmom}
\end{equation}
where $\epsilon ^{ij}$, $i,j=1,2$, is the two dimensional totally
antisymmetric tensor satisfying the definition $\epsilon ^{12}=1$. A
straightforward calculation shows that the C--S Hamiltonian is given by: 
\begin{equation}
H_{CS}=-\int d^2{\bf x}A_0^a\left( D_i^{ab}\pi ^{i,b}+\partial _i\pi
^{i,a}\right) \label{wrrr}
\end{equation}
In the above equation $D_i^{ab}$ denotes the spatial components of the
covariant derivative: 
\[
D_\mu ^{ab}\equiv \partial _\mu \delta ^{ab}+f^{abc}A_\mu ^c
\]
From eqs. (\ref{canmom}) we obtain the constraints:
\begin{eqnarray}
\varphi ^{0,a} &=&\pi ^{0,a}  \label{fpconstr} \\
\varphi ^{i,a} &=&\pi ^{i,a}-\frac s{8\pi }\epsilon ^{ij}A_j^a\qquad \qquad
\qquad i=1,2  \label{spconstr}
\end{eqnarray}
Following the Dirac procedure for constrained systems,
the latter will be imposed in the weak sense:
$$\varphi^{\mu,a}\approx 0$$
To this purpose, we
construct the
extended Hamiltonian: 
\begin{equation}
\widetilde{H}=H_{CS}+\int \lambda _\mu ^a\varphi ^{\mu ,a}d^2{\bf x}
\label{extham}
\end{equation}
where the $\lambda _\mu ^a$'s represent the
Lagrange multipliers corresponding to the
primary constraints $\varphi^{\mu,a}$.

From the consistency conditions
$\dot \varphi
^{\mu ,a}=\left\{ \varphi ^{\mu ,a},\widetilde{H}_{CS}\right\} \approx 0$,
we obtain the secondary constraint:
\newfont\prova{eusm10 scaled\magstep1}
\begin{equation}
\text{{\prova G}}^a=
D_i^{ab}\pi ^{i,b}+\partial _i\pi
^{i,b}\approx 0 \qquad\qquad\qquad \text{Gauss law} \label{glaw}
\end{equation}
and two relations which  determine the Lagrange multipliers $\lambda_1$ and
$\lambda_2$:
\begin{equation}
\frac s{4\pi }\epsilon ^{ij}\left( D_j^{ab}A_0^b-\lambda _j^a\right)\approx
0\qquad \qquad \qquad i=1,2  \label{lagdet}
\end{equation}
It is possible to see that the
consistency condition  \.{\prova G}$^a\approx 0$ does not lead to any
further independent equation.
Let us notice that
the operators $\text{\prova G}^a$ generate the $SU(N)$ group of gauge
transformations. To show this,
we introduce the
Dirac brackets (DB's)
associated to
the second class
constraints $\varphi _i^a$ of eq. (\ref{spconstr}): 
\begin{equation}
\left\{ A({\bf x}),B({\bf y})\right\} ^{*}=\left\{ A({\bf x}),B({\bf y}%
)\right\} -\sum_{i,j=1}^2\int d^2{\bf x}^{\prime }d^2{\bf y}^{\prime
}\left\{ A({\bf x}),\varphi ^i({\bf x}^{\prime })\right\} \left(
C^{-1})_{ij}({\bf x}^{\prime },{\bf y}^{\prime }\right) \left\{ \varphi ^j(%
{\bf y}^{\prime }),B({\bf y})\right\}   \label{dbdef}
\end{equation}
where $\left( C^{-1}\right) _{ij}({\bf x},{\bf y})$ is the inverse of the
matrix $C^{ij}({\bf x},{\bf y})=\left\{ \varphi ^i({\bf x}),\varphi ^j({\bf y%
})\right\} $. For simplicity, color indices and the time variable have been
omitted in the above equations.\\ Computing $\left( C^{-1}\right) _{ij}({\bf %
x},{\bf y})$ explicitly, we find: 
\[
\left( C^{-1}\right) _{ij}^{ab}({\bf x},{\bf y})=\frac{4\pi }s\delta
^{ab}\epsilon _{ij}\delta ({\bf x}-{\bf y})
\]
The Dirac brackets among the canonical variables are now given by 
\begin{eqnarray}
\left\{ A_i^a(t,{\bf x}),\pi ^{j,b}(t,{\bf y})\right\} ^{*} &=&\frac 12%
\delta ^{ab}\delta _i^j\delta ({\bf x}-{\bf y})  \label{bone} \\
\left\{ A_i^a(t,{\bf x}),A_j^b(t,{\bf y})\right\} ^{*} &=&\frac{4\pi }s%
\delta ^{ab}\epsilon _{ij}\delta ({\bf x}-{\bf y})  \label{btwo} \\
\left\{ \pi ^{i,a}(t,{\bf x}),\pi ^{j,b}(t,{\bf y})\right\} ^{*} &=&\frac s{%
16\pi }\delta ^{ab}\epsilon ^{ij}\delta ({\bf x}-{\bf y})  \label{bthree}
\end{eqnarray}
In the following, the DB's defined in
(\ref{dbdef}) will be written without the
superscript $^*$.
Exploiting the DB's (\ref{bone})--(\ref{bthree}), we
obtain the relations:
\begin{eqnarray}
\left\{ \text{{\prova G}}^a(t,{\bf x}),A_i^b(t,{\bf y})\right\} 
&=&-D_i^{ab}(x)\delta ({\bf x}-{\bf y})\label{ffcone} \\
\left\{ \text{{\prova G}}[\psi ],A_i^a(t,{\bf x})\right\} 
&=& D_i^{ab}(x)\psi ^b({\bf x})\label{ffctwo} \\
\left\{ \text{{\prova G}}^a(t,{\bf x}),\text{{\prova G}}^b(t,{\bf y}%
)\right\}  &=&-f^{abc}\text{{\prova G}}^c(t,{\bf x})\delta ({\bf x-y)}
\label{ffcthree}
\end{eqnarray}
where $\text{\prova G}\left[ \psi \right] =\int d^2{\bf x}{\prova G}^a(t,{\bf
x})\psi ^a({\bf x})$.
This shows that the $\text{\prova G}^a(t,{\bf x})$ are the generators of the $%
SU(N)$ gauge transformations as desired.

At this point, we are left with the constraints given by eq.
(\ref{fpconstr}) and by the Gauss law
(\ref{glaw}). 
However, the former constraint, which involves the conjugate momentum of
$A_0^a$ can be ignored.
As a matter of fact, the field
$A_0^a$ plays just
the role of the Lagrange multiplier associated to the Gauss law
in the Hamiltonian (\ref{wrrr}) and has no dynamics.
From eqs. (\ref{ffcone})--(\ref{ffcthree}) it turns out that the Gauss law
(\ref{glaw}) is
a first class constraint. To make it second class,
we introduce the Coulomb gauge
fixing:
\begin{equation}
\partial _iA^{i,a}\approx 0  \label{coulombgauge}
\end{equation}
and the new extended Hamiltonian:
\begin{equation}
\check H_{CS}=\int d^2{\bf x}\left[ -A_0^a\text{{\prova G}}^a+\frac s{8\pi }%
A_i^a\partial ^iB^a+\lambda _0^a\pi ^{0,a}\right] \label{neham}
\end{equation}
From the condition  $\{\partial_i A^{i,a},\check H_{CS}\}
\approx 0$, we obtain an equation for $A_0^a$:
\begin{equation}
\partial^iD_i^{ab}A_0^b\approx 0\label{sfour}
\end{equation}
Moreover, the requirement
$\left\{ \partial^iD_i^{ab}A_0^b(x),\check H_{CS}
\right\}\approx 0$ determines the
Lagrange multiplier $\lambda_0$:
\begin{equation}
-\bigtriangleup\lambda_0^a
-\left\{\partial_i({\bf A}_i\times{\bf A}_0)^a,\check H_{CS}\right\}\approx0
\label{lzdet}
\end{equation}
In the above equation the symbol $\bigtriangleup$ denotes the two  dimensional
Laplacian $\bigtriangleup=-\partial_i\partial^i$ and
$$({\bf A}_i\times{\bf A}_0)^a\equiv f^{abc}A_i^bA_0^c$$
Another independent equation, which fixes the Lagrange multipliers $B^a$, is
provided by the requirement \.{\prova G}$^a\approx 0$:
\begin{equation}
\left\{{\text{\prova G}}^a,\check H_{CS}\right\}
\approx-\frac{s}{8\pi}D_i^{ab}\partial^i
B^b\approx 0\label{fixb}
\end{equation}
Let us notice that the above relations (\ref{glaw}), (\ref{coulombgauge}) and
(\ref{sfour})--(\ref{fixb}) are compatible with the equations of motion of the
gauge potentials:
\begin{equation} \epsilon^{ij}(D_i^{ab}A_j^{b}-\partial_jA_i^a)=0\label{eqone}
\end{equation}
\begin{equation} D_j^{ab}A_0^b-\partial_0A_j^a=0\label{eqtwo}
\end{equation}
As a matter of fact (\ref{eqone}) is equivalent to the condition
${\text{\prova G}}^a=0$. Moreover, multiplying for instance eq. (\ref{eqtwo})
with the differential operator $\epsilon_{ki}\partial^k$, we obtain the
relation:
$$\partial_0\partial^kA_k^a-\partial^kD_k^{ab}A_0^b=0$$
which is consistent with the Coulomb gauge
and the condition (\ref{sfour}) on $A_0^a$.

It is now possible to realize that the Gauss law
(\ref{glaw}) and the Coulomb gauge fixing
(\ref{coulombgauge}) form a set of second class constraints, so that we can
impose them
in the strong sense computing the final Dirac brackets.
Putting
\[
\chi_1^a=\text{\prova G}^a\qquad\qquad\qquad
\chi_2^a=
\partial_i A^{i,a}
\]
with $\alpha,\beta=1,2$, we have for
any two observables $A({\bf x})$ and
$B({\bf y})$ \footnotemark\footnotetext{In the following, the time
variable will be omitted from our equations.}:
\[
\left\{ A^a({\bf x}),B^b({\bf y})\right\}^{*}=\left\{ A^a({\bf x}),B^b({\bf y}%
)\right\}-
\]
\begin{equation}
\sum_{\alpha,\beta=1}^2\sum_{c,d}
\int d^2{\bf x}^{\prime }d^2{\bf y}^{\prime
}\left\{ A^a({\bf x}),\chi_\alpha^c({\bf x}^{\prime })\right\}(
C^{-1})^{\alpha\beta,cd}({\bf x}^{\prime },{\bf y}^{\prime })
\left\{ \chi ^{\beta,d}(%
{\bf y}^{\prime }),B^b({\bf y})\right\}   \label{dbndef}
\end{equation}
The matrix 
$(C^{-1})^{\alpha\beta,cd}({\bf x},{\bf y})$ denotes the inverse of the
$2\times 2$ matrix $C_{\alpha\beta}^{ab}({\bf x},{\bf y})=\{\chi^a_\alpha
({\bf x}),\chi^b_\beta({\bf y})\}$.
After some manipulations and remembering that the gauge potentials
satisfy the Coulomb gauge constraint, we obtain:
\[
{\bf C}^{ab}({\bf x},{\bf y})=\left(
\begin{array}{c c }
0 & -D^{ab}_i({\bf x})\partial_{\bf x}^i\delta({\bf x}-{\bf y})\\
D^{ab}_i({\bf x})\partial_{\bf x}^i\delta({\bf x}-{\bf y})  & 0\\
\end{array}\right)
\]
To invert the above matrix, it
is convenient to introduce the function $\text{\prova D}^{cb}
({\bf x},{\bf y})$, defined by the following equation \cite{schwinger}:
\begin{equation}
D^{ac}_i({\bf x})\partial_{\bf x}^i\text{\prova D}^{cb}({\bf x},{\bf y})=
\delta^{ab}\delta({\bf x}-{\bf y})\label{dstorta}
\end{equation}
Supposing that
the Green function $\text{\prova D}^{ab}({\bf x},{\bf y})$ has a sufficiently
good behavior at infinity, it is easy to prove that
\begin{equation}
({\bf C}^{-1})^{ab}({\bf x},{\bf y})=\left(
\begin{array}{c c }
0 & \text{\prova D}^{ab}({\bf x},{\bf y})\\
-\text{\prova D}^{ab}({\bf x},{\bf y})  & 0\\
\end{array}\right)\label{invnew}
\end{equation}
After imposing the constraints (\ref{glaw}) and (\ref{coulombgauge}) in the
strong sense, the Hamiltonian $\check H_{CS}$ vanishes, but the commutation
relations (CR's) between the fields remain complicated.
From eqs. (\ref{dbndef}) and (\ref{invnew}), in fact,  the basic DB's
between the canonical variables $A_i^a$ have the following form:
\[
\left\{A_i^a({\bf x}),A_j^b({\bf y})\right\}^*=
-\frac{4\pi}{s}\delta^{ab}\epsilon_{ij}\delta({\bf x}-{\bf y})+
\]
\begin{equation}
\frac{4\pi}{s}\epsilon_{ik}\partial_{\bf x}^kD_j^{bc}({\bf y})
\text{\prova D}^{ac}
({\bf x},{\bf y})-\frac{4\pi}{s}
\epsilon_{kj}D_i^{ac}({\bf x})\partial_{\bf y}^k\text{\prova D}^{cb}
({\bf x},{\bf y})\label{maincomrel}
\end{equation}
Let us study the main properties of the above DB's. First of all, they
are antisymmetric as expected:
\begin{equation}
\{ A_i^a({\bf x}), A_j^b({\bf y})\}^*=-
\{ A_j^b({\bf y}), A_i^a({\bf x})\}^*\label{antisym}
\end{equation}
The antisymmetry of the right hand side of eq. (\ref{maincomrel}) is not
explicit, but can be verified with the help of the relation:
\begin{equation}
\label{propsym}
\text{\prova D}^{ab}({\bf x},{\bf y})=\text{\prova D}^{ba}({\bf y},{\bf x})
\end{equation}
The above symmetry of the Green function $\text{\prova D}^{ab}({\bf x},{\bf
y})$ in its arguments is a consequence of the selfadjointness of the
defining equation (\ref{dstorta}) \cite{schwinger}.
Moreover, the CR's (\ref{maincomrel}) are consistent
with the Coulomb gauge. As a matter of fact, it is easy to prove that:
$$\{ A_i^a({\bf x}), \partial^jA_j^b({\bf y})\}^*=
\{ \partial^iA_i^a({\bf x}), A_j^b({\bf y})\}^*=0$$
The case of a Chern--Simons field theory with abelian gauge group $U(1)$ is
particularly instructive in order to understand the meaning of the CR's
(\ref{maincomrel}). Let $U_\mu$ denote the abelian gauge fields.
Then the Lagrangian (\ref{lagrangian}) reads:
$$L_{CS}={s\over 8\pi}\epsilon^{\mu\nu\rho}U_\mu\partial_\nu U_\rho$$
It is now possible to decompose the gauge potentials $U_i$, $i=1,2$ into
transverse and longitudinal components:
$$U_i=\epsilon^{ij}\partial_j\varphi+\partial_i \rho$$
where $\varphi$ and $\rho$ are two real scalar fields.
Exploiting the Coulomb gauge condition it turns out that $\rho=0$.
The canonical momenta are given by:
$$\pi^i=\frac{s}{8\pi}\epsilon^{ij}U_j$$
As a consequence,
from the Gauss law $\partial_i\pi^i=0$, we obtain the relation $\partial_i
\partial^i\varphi=0$. This implies that $\varphi=0$ and thus
there is no dynamics in the C--S
field theory as expected.

The CR's (\ref{maincomrel}) must be consistent with that fact.
Indeed, in the abelian case it is easy to derive the Green function
$\text{\prova D}({\bf x},{\bf y})$ solving eq. (\ref{dstorta}). The result is:
\begin{equation}
\text{\prova D}({\bf x},{\bf y})=-{1\over 2\pi} {\rm log}|
{\bf x}-{\bf y}|\label{dsabelian}
\end{equation}
Substituting the right hand side of the above equation in (\ref{maincomrel}),
we obtain:
$$[U_i(t,{\bf x}),U_j(t,{\bf y})]=0$$
so that the fields do not propagate as expected.
To conclude the discussion of the abelian case, let us notice that
eqs. (\ref{lagdet}) and (\ref{sfour})--(\ref{fixb}) admit only the
trivial solutions $U_0=\lambda_\mu=B=0$ in agreement with the fact that,
in absence of couplings with matter fields,
the C--S theory is topological and there
are no degrees of freedom. \smallskip
In the nonabelian case the situation is analogous, but the equations
of motion of the constraints become nonlinear and can in general be solved
only using a perturbative approach. The relevant equations determining the
fields $A_i^a(z)$, with $i=1,2$, are given by:
\begin{equation}
F_{12}^a=\partial_1A_2^a-\partial_2A_1^a-gf^{abc}A_1^bA_2^c\label{glclone}
\end{equation}
and 
\begin{equation}
\partial_1A_1^a-\partial_2^aA_2^a=0\label{cgclone}
\end{equation}
With respect to eq. (\ref{lagrangian}),
we have introduced here the new coupling
constant $g^2=\frac{8\pi}{9s}$ and the fields $A_\mu$ have been rescaled
in such a way that the new action becomes:
$$
L=\epsilon^{\mu \nu \rho }\left( A_\mu ^a\partial _\nu A_\rho
^a-gf^{abc}A_\mu ^aA_\nu ^bA_\rho ^c\right)
$$
In the following, we will also suppose that $g$ is so small that
a perturbative treatment of the C--S field theory makes sense.
Under this hypothesis, the fields $A_i^a$ can be expanded in powers of $g$:
$$A_i^a(x)=\sum_{n=0}^\infty g^nA_i^{a (n)}(x)$$
where, from eqs. (\ref{glclone}) and (\ref{cgclone}), the $A_i^{a (n)}$'s
satisfy the following equations:
$$\partial_1A_2^{a(0)}-\partial_2A_1^{a(0)}=0\qquad\qquad\qquad
\partial_1A_1^{a(0)}+\partial_2A_2^{a(0)}=0$$
and
\begin{equation}
\partial_1A_2^{a(n)}-\partial_2A_1^{a(n)}-gf^{abc}
A_1^{b(n-1)}A_2^{c(n-1)}=0\qquad\qquad\qquad n=1,\ldots,\infty
\label{hoone}
\end{equation}
\begin{equation}
\partial_1A_1^{a(n)}+\partial_2A_2^{a(n)}=0
\qquad\qquad\qquad n=1,\ldots,\infty
\label{hotwo}
\end{equation}
Assuming that the gauge fields vanish at infinity, the solution of the above
equations at the zeroth order is
\begin{equation}
A_1^{a(0)}(t,{\bf x})=A_2^{a(0)}(t,{\bf x})=0\label{cvd}
\end{equation}
as shown in the abelian case.
Moreover, from eq. (\ref{hoone}), it turns out that $A_1^{a(n)}(t,{\bf x})=0$
for $n=1,\ldots,\infty$, so that all the field configurations solving
eqs. (\ref{glclone})--(\ref{cgclone}) vanish identically.
Pure gauge solutions obtained performing gauge transformations
are not allowed because, at least within
perturbation theory, the Coulomb gauge fixes the gauge freedom completely.
As a consequence, the right hand side
of (\ref{maincomrel}) is equal to zero. Indeed, due to eq. (\ref{cvd}), the
Green function $\text{\prova D}^{ab}({\bf x},{\bf y})$
is given by:
\begin{equation}
\text{\prova D}^{ab}({\bf x},{\bf y})=-\delta^{ab}{1\over 2\pi} {\rm log}|
{\bf x}-{\bf y}|\label{dab}
\end{equation}
and, substituting in eq. (\ref{maincomrel}), we obtain:
\begin{equation}
\{ A_i^a({\bf x}), A_j^b({\bf y})\}^*=0\label{physbra}
\end{equation}
as expected.\\
Of course, the vanishing of the gauge fields leads to the
trivial solutions $A_0^a=\lambda_\mu^a=B^a=0$  for the Lagrange multipliers
as in the abelian case.\\
It is worth remarking,  that the would be Poincar\'e
algebra becomes trivial {\it a posteriori}, that is
when computed on the "physical" solutions 
of the theory (eq. \ref{cvd}), characterized by the "strong" validity of the
constraints and of the brackets given in eq. (\ref{physbra}).
That means that the Poincar\'e covariance is recovered through the
trivial representation of the Poincar\'e group \footnotemark{} \footnotetext{
It is worth mentioning that in the case of the Maxwell--Chern--Simons
theory the Poincar\'e covariance takes place through a {\bf nontrivial}
representation of the Poincar\'e group \cite{gg}.}.
We stress the fact that one must evaluate in such {\it a posteriori} way the
algebra, as otherwise one finds "extra" terms, proportional to the
constraints.  For instance, the intermediate Dirac brackets (
\ref{bone}--\ref{bthree}) yield for
the generators of the time and the space translations the following result:
$$\{P_0, P_k\} = \int d^2x A_0^a \partial_k \text{\prova G}^a$$
where $\text{\prova G}$ is given in (\ref{glaw}).
Let us notice that in the case of the MCS theory the Poncar\'e invariance
has been proved in the Coulomb gauge within the frame of the canonical
formalism in ref. \cite{gg}.

To quantize the theory, we have to replace
the Dirac brackets (\ref{maincomrel}) with commutators.
At least in the absence of coupling with matter fields, we obtain
trivial commutation relations between the gauge potentials:
\begin{equation}
\left[A_i^a({\bf x}),A_j^b({\bf y})\right]=0\label{quantumcr}
\end{equation}

\section{Conclusions}
In this paper the C--S field theories have been quantized in the Coulomb gauge
within the Dirac's canonical approach to constrained systems.
All the constraints coming from the Hamiltonian procedure and by the
Dirac's consistency requirements have been derived.
As anticipated in the Introduction, the C--S theories become
in this gauge two dimensional
models. Only the fields $A_i^a$, for $i=1,2$, have in fact a dynamics, which
is governed by the commutation relations (\ref{maincomrel}).
If no interactions with matter fields are present, we have shown
that these CR's vanish at all perturbative orders.
Thus the C--S field theories in the Coulomb gauge are not only finite,
but also free.
This result has been verified with explicit perturbative
calculations of the correlation functions in \cite{flnew}.
A natural question that arises at this point is if analogous conclusions
can be drawn for the covariant gauges.
For this reason it would be
interesting to repeat the procedure of canonical quantization developed
here 
also in this case.

The situation becomes different
if the interactions with other fields are switched on.
Adding for instance a coupling with a current $J^a_\mu$ of the kind
$\int d^2{\bf x}A_\mu^a J^{\mu,a}$ to the Hamiltonian (\ref{neham}),
it is possible to see that the Gauss law (\ref{glaw}) is modified as follows:
$$D_i^{ab}\pi ^{i,b}+\partial _i\pi
^{i,b}+J_0^a\approx 0$$
Thus eqs. (\ref{cvd}) are no longer valid and we have to consider the
full commutation relations (\ref{maincomrel}).
Remarkably,
they trivially vanish at the zeroth level in the coupling constant $g$.
Moreover, the CR's (\ref{maincomrel})
are perfectly well defined and do not lead
to ambiguities in the quantization of the
C--S models in the Coulomb gauge.
In particular, we have verified here the consistency of (\ref{maincomrel})
with the Coulomb gauge fixing and their antisymmetry under the
exchange of the fields.

A physical application of our results,
which is currently under consideration, is the
investigation of the statistics of fermionic and bosonic matter
fields interacting with nonabelian C--S theories at high temperatures
\cite{higtemp}. Other interesting applications are $(2+1)$
quantum gravity and the calculation of the new link invariants
from C--S field theories quantized
on Riemann surfaces, whose existence has been formally shown
in \cite{cotta}. In these latter two cases,
the possibility offered by the Coulomb gauge of performing explicit
calculations also on non--flat space--times \cite{ffprd}
can be exploited.


\begin{thebibliography}{99}
\bibitem{csft} R. Jackiw and S. Templeton, {\it Phys. Rev.} {\bf D23} (1981),
2291; S. Deser, R. Jackiw and S. Templeton, {\it Phys. Rev. Lett.}
{\bf 48} (1983), 975.
\bibitem{csftt} 
C. Hagen, {\it ibid.} {\bf 157} (1984), 342; {\it Phys. Rev.} {\bf D31} (1985),
2135; J. Schonfeld, {\it Nucl. Phys.} {\bf B185} (1981), 157.
\bibitem{witten} E. Witten, {\it Comm. Math. Phys.} {\bf 121} (1989), 351. 
\bibitem{dirac} P. A. M. Dirac, {\it Lectures in Quantum Mechanics},
Yeshiva University Press, New York 1964.
\bibitem{hrt} A. Hanson, T. Regge and C. Teitelboim, {\it Constrained
Hamiltonian Systems}, Accademia dei Lincei, Roma, 1976 and references
therein.
\bibitem{ffprd} F. Ferrari, {\it Phys. Rev.} {\bf D50} (1994), 7578.
\bibitem{csappl} G. Moore and N. Seiberg, {\it Phys. Lett.} {\bf B220} (1989),
422; J. Fr\"ohlich and C. King, {\it Comm. Math. Phys.} {\bf 126}
(1989), 167;
E. Guadagnini, M. Martellini and M. Mintchev, {\it Nucl. Phys.}
{\bf B336} (1990), 581; 
G. W. Semenoff, {\it Phys. Rev. Lett} {\bf 61} (1988), 517; E. Fradkin {\it
Phys. Rev. Lett.} {\bf 63} (1989), 322; M. L\"uscher, {\it Nucl. Phys.}
{\bf B326} (1989), 557; 
E. Witten, {\it Nucl. Phys.} {\bf B311} (1988), 46; 
Y. H. Chen, F. Wilczek, E. Witten and B. I. Halperin, {\it Int. Jour. Mod.
Phys.} {\bf B3} (1989), 1001;
R. Iengo and K. Lechner, {\it Phys. Rep.} {\bf 213} (1992), 179.
\bibitem{cgincs} S. Deser, R. Jackiw and S. Templeton,
{\it Ann. Phys.} (N. Y.) {\bf 140} (1984), 372;
A. Bellini, M. Ciafaloni and P. Valtancoli, {\it Nucl. Phys.}
{\bf B454} (1995), 449; {\bf B462} (1996), 453; {\it Phys. Lett.};
K. Haller and E. L. Lombridas, {\it Ann. Phys.} {\bf 246} (1996), 1.
\bibitem{gg} F. P. Devecchi, M. Fleck, H. O. Girotti, M. Gomes and A. J. da
Silva, {\it Ann. Phys.} {\bf 242} (1995), 275; O. Bergman and G. Lozano,
{\it Ann. Phys.} (NY) {\bf 229} (1994), 229; D. Bak and O. Bergman, {\it Phys.
Rev.} {\bf D51} (1995), 1994; M.-I. Park and Y.-J. Park, {\it Phys. Rev.}
{\bf D50} (1994), 7584.
\bibitem{hh} A. Foerster and H. O. Girotti, {\it Statistical transmutations
in $2+1$ dimensions}, in J. J. Giambiagi Festschrift, eds. H. Falomir, R. E.
Gamboa Saravi, P. Leal Ferreira and F. A. Schaposnik (World Scientific,
Singapore, 1990), p. 161; {\it Phys. Lett.} {\bf B230} (1989), 83;
{\it Nucl. Phys.} {\bf B342} (1990), 680.
\bibitem{taylor} P. J. Doust and J. C. Taylor, {\it Phys. Lett.}
{\bf 197B} (1987), 232; P. J. Doust, {\it Ann. Phys.} (N. Y.) {\bf 177} (1987),
169.
\bibitem{leibbrandt} G. Leibbrandt, {\it Noncovariant Gauges}, World
Scientific, Singapore, 1994.
\bibitem{chetsa} H. Cheng and E. C. Tsai, {\it Phys. Rev. Lett.}
{\bf 57} (1986), 511.
\bibitem{covgau} L. Alvarez--Gaum\'e, J. M. F. Labastida and A. V. Ramallo,
{\it Nucl. Phys.} {\bf B334} (1990), 103;
W. Chen, G. W. Semenoff and Y. S. Wu, {\it Mod. Phys. Lett.} {\bf A5}
(1990), 1833 {\it Phys. Rev.} {\bf D46} (1992), 5521;
D. Birmingham, M. Rakowsky and G. Thompson, {\it Phys. Lett.}
{\bf B251} (1990), 121; E. Guadagnini, M. Martellini and M. Mintchev,
{\it Phys. Lett.} {\bf B227} (1989), 111;
A. Blasi and R. Collina, {\it Nucl. Phys.} {\bf B345} (1990), 472;
M. Asorey and F. Falceto, {\it Phys. Lett.} {\bf B241} (1990), 31;
F. Delduc, C. Lucchesi, O. Piguet and S. P. Sorella,
{\it Nucl. Phys.} {\bf B346} (1990), 313.
\bibitem{schwinger} J. Schwinger, {\it Phys. Rev.} {\bf 125} (1962), 1043.
\bibitem{linni} Q.-G. Lin and G.-J. Ni, {\it Class. Quantum Grav.} {\bf 7}
(1990), 1261.
\bibitem{higtemp} G. Dunne, R. Jackiw, S.-Y. Pi and C. A. Trugenberger,
{\it Phys. Rev.} D {\bf 43} (1991), 1332; G. Dunne, {\it Comm. Math. Phys.}
{\bf 150} (1993, 519.
\bibitem{cotta} P. Cotta Ramusino, E. Guadagnini, M. Martellini and
M. Mintchev, {\it Nucl. Phys.} {\bf B330} (1990), 557.
\bibitem{flnew} F. Ferrari and I. Lazzizzera, {\it Perturbative
Analysis of the Chern--Simons Field Theory in the Coulomb Gauge},
Preprint UTF 387/96, PAR-LPTHE 96-44.
\end{thebibliography}
\end{document}